\documentstyle[aps,multicol,prl,epsf]{revtex}

\begin{document}

\draft 
\twocolumn[\hsize\textwidth\columnwidth\hsize\csname@twocolumnfalse%
\endcsname

\title{Inclusive Electron-Nucleus Scattering at Large Momentum Transfer}
\author{J. Arrington$^{2*}$, 
C. S. Armstrong$^{11\#}$, 
T. Averett$^{2+}$, 
O. K. Baker$^{4,9}$, 
L. de Bever$^{1}$, 
C. W. Bochna$^{5}$, 
W. Boeglin$^{3}$, 
B. Bray$^{2}$, 
R. D. Carlini$^{9}$, 
G. Collins$^{6}$, 
C. Cothran$^{10}$, 
D. Crabb$^{10}$, 
D. Day$^{10}$, 
J. A. Dunne$^{9\&}$, 
D. Dutta$^{7}$, 
R. Ent$^{9}$, 
B. W. Filippone$^{2}$, 
A. Honegger$^{1}$, 
E. W. Hughes$^{2}$, 
J. Jensen$^{2}$, 
J. Jourdan$^{1}$, 
C. E. Keppel$^{4,9}$, 
D. M. Koltenuk$^{8}$, 
R. Lindgren$^{10}$, 
A. Lung$^{6\#}$, 
D. J Mack$^{9}$, 
J. McCarthy$^{10}$, 
R. D. McKeown$^{2}$, 
D. Meekins$^{11}$, 
J. H. Mitchell$^{9}$, 
H. G. Mkrtchyan$^{9}$, 
G. Niculescu$^{4}$, 
I. Niculescu$^{4}$, 
T. Petitjean$^{1}$, 
O. Rondon$^{10}$, 
I. Sick$^{1}$, 
C. Smith$^{10}$, 
B. Terburg$^{5}$  
W. F. Vulcan$^{9}$, 
S. A. Wood$^{9}$, 
C. Yan$^{9}$, 
J. Zhao$^{1}$, 
B. Zihlmann$^{10}$ }

\address {$^{1}$University of Basel, Basel Switzerland\\
$^{2}$Kellogg Radiation Laboratory, California Institute of Technology, Pasadena CA 91125\\
$^{3}$Florida International University, University Park FL 33199\\
$^{4}$Hampton University, Hampton VA 23668\\
$^{5}$University of Illinois, Urbana-Champaign IL 61801\\
$^{6}$University of Maryland, College Park MD 20742\\
$^{7}$Northwestern University, Evanston IL 60201\\
$^{8}$University of Pennsylvania, Philadelphia PA 19104\\
$^{9}$Thomas Jefferson National Accelerator Facility, Newport News VA 23606\\
$^{10}$University of Virginia, Charlottesville VA 22901\\
$^{11}$College of William and Mary, Williamsburg, VA 23187\\
} 

\date{Nov. 10, 1998}
\maketitle

\begin{abstract}
Abstract: Inclusive electron scattering is measured with 4.045 GeV incident 
beam energy from C, Fe and Au targets. The measured energy transfers and 
angles correspond to a kinematic range for Bjorken $x > 1$ and momentum 
transfers from $Q^2 = 1 - 7$ (GeV/c)$^2$. When analyzed in terms of the 
$y$-scaling function the data show for the first time an approach to scaling 
for values of the initial nucleon momenta significantly greater than the 
nuclear matter Fermi-momentum (i.e. $> 0.3$ GeV/c). 

\medskip
\end{abstract}
]


High energy electron scattering from nuclei can provide important information 
on the wave function of nucleons in the nucleus. In particular, with simple 
assumptions about the reaction mechanism, scaling functions can be 
deduced that, 
if shown to scale (i.e. are independent of length scale or momentum 
transfer), can provide information about the momentum and energy distribution 
of nucleons in a nucleus. Several theoretical 
studies~\cite{annrev,frank,cdasim1,ben1} have indicated 
that such measurements may provide direct access to short-range 
nucleon-nucleon correlations. 

The concept of y-scaling in electron-nucleus scattering was first introduced 
by West~\cite{west} and Kawazoe et al.~\cite{kawa}. 
They showed that in the impulse approximation, if quasielastic 
scattering from a nucleon in the nucleus was the dominant reaction mechanism, 
a scaling function $F(y)$ could be extracted from the measured cross section 
which was related to the momentum distribution of the nucleons in the nucleus.
In the simplest approximation the corresponding scaling variable $y$  is 
the minimum momentum of the struck nucleon along the direction of the virtual 
photon. In general the scaling function depends on both $y$ and momentum 
transfer - $F(y,Q^2)$ - but at sufficiently high $Q^2$ ($-Q^2$ is the 
square of the 
four-momentum transfer) the dependence on $Q^2$ should vanish yielding 
scaling. 
However the simple impulse approximation picture breaks down when the 
final-state interactions (FSI) of the struck nucleon with the
rest of the nucleus are included. Previous 
calculations~\cite{fsi1,fsi2,ben2,fsi3,fsi4,fsi5,fsi6,fsi8} suggest that the
contributions from final state interactions should vanish at sufficiently 
high $Q^2$. A previous SLAC measurement~\cite{ne3} suggested 
an approach to the scaling limit for heavy nuclei but only for low values of 
$|y| < 0.3$ GeV/c at momentum transfers up to 3 (GeV/c)$^2$. The data 
presented here 
represent a significant increase in the Q$^2$ range compared to previous
measurements while also extending the coverage in $y$. 

The present data were obtained in Hall C at the Thomas Jefferson National 
Accelerator Facility (TJNAF), using 
4.045 GeV electron beams with intensities from 10 - 80 $\mu$A.  The absolute 
beam energy was calibrated to 0.08\% using 0.8 GeV elastic scattering from 
carbon and BeO targets and 4.0 GeV elastic scattering from
hydrogen. The beam current was monitored with three calibrated resonant 
cavities. The beam energy 
resolution was better than 0.05\% as defined by the accelerator acceptance. 
Solid targets of C (2.1 and 5.9\% of a radiation length), Fe 
(1.5 and 5.8\% of a radiation length) and Au (5.8\% of a 
radiation length) with natural isotopic abundance were used. Data were also 
taken with liquid targets of hydrogen and deuterium (nominally 4 and 15 cm 
in length). Scattering from hydrogen 
allows a cross check of the absolute normalization of the cross section; 
results from the deuterium target 
will be presented elsewhere. Less than 1\% density 
variations were observed for the liquid targets due to beam heating 
for incident beam currents up to 55 
$\mu$A (maximum current used for the liquid targets) 
when the 200 $\mu$m $\times$ 200 $\mu$m beam was rastered by a pair
of electro-magnets to the typical spot-size of $\pm$ 1.2 mm. 

The scattered electrons were detected with the High Momentum Spectrometer 
(HMS) at angles of $15^\circ, 23^\circ, 30^\circ, 37^\circ, 45^\circ$ and 
$55^\circ$ and the Short Orbit Spectrometer (SOS) at an angle of $74^\circ$. 
Both spectrometers took data simultaneously with 
nearly identical detector systems configured for electron detection. Each
detector system included two planes of plastic scintillator for triggering, 
two six-element drift chambers for tracking information as well as a 
gas \v{C}erenkov detector and Pb glass calorimeter for particle 
identification. 

The measured tracks were required to reconstruct to the target location.
For the HMS, additional cuts were 
applied to eliminate events produced on the pole pieces of the spectrometer
magnets. Cuts were also applied to select electrons and reject $\pi^-$ 
using the signals from the Cerenkov detector and Calorimeter. The 
combined efficiency of all
the cuts was $> 98\%$. The binned events were corrected for spectrometer 
acceptance using an acceptance function generated by a Monte Carlo 
calculation~\cite{jra} that
included all apertures within the spectrometer. This calculation accurately 
reproduced the distributions and cross section from hydrogen elastic 
scattering.
Estimated systematic uncertainties due to the acceptance are $< 2.5\%$. 
Tracking efficiencies were typically 94\% - 97\%. Background from 
mis-identified $\pi^-$ was negligible for the HMS and $< 3\%$ in the 
worst case for the SOS. High energy photons produced principally from 
$\pi^0$ decay can result in secondary electrons following pair production 
by the 
photons in the target material. This background, estimated by 
measuring positron yields with the spectrometer magnetic fields reversed, was 
negligible for spectrometer angles $< 55^\circ$, but was 3 - 10\% at
$55^\circ$ and 20 - 100\% at $74^\circ$. The larger values for the 
contribution of 
this background are for the 6\% radiation length targets and result
in an estimated
systematic error of 5 - 10\%. However, because the large backgrounds are only 
present in kinematic regions where the cross section is very small, the 
statistical uncertainties dominate the total uncertainty. 

Because of the large acceptance of the spectrometers ($>6$ msr) and the 
rapid variation of the cross section with $\theta$, there can 
be a significant variation of the cross section over the acceptance. 
In order to extract cross sections vs. energy transfer $\nu$ at fixed 
scattering
angle a bin centering correction must be applied. This is accomplished with 
a model of the cross section~\cite{jra} 
that is constrained to reproduce the angle
and energy transfer dependence of the measurements. 
The cross section model was also 
used to apply radiative corrections using the iterative technique of 
Refs.~\cite{stein} and ~\cite{dhp}. 
Variations in the form of the model were used to estimate systematic 
uncertainties in these corrections. The total estimated systematic 
uncertainties
in the bin-centering and radiative corrections were 1-2\% and 2.5\% 
respectively.
Lastly a Coulomb correction was applied for the change in the incident and
scattered energy due to the Coulomb acceleration from the nuclear charge. 
This correction was significant ($\sim 10\%$ for Fe and $\sim 20\%$ for Au) 
for the largest scattering angles of the present experiment. 
 
\begin{figure}[sigma]
\epsfxsize 8.5 cm \epsfysize 6 cm \epsfbox{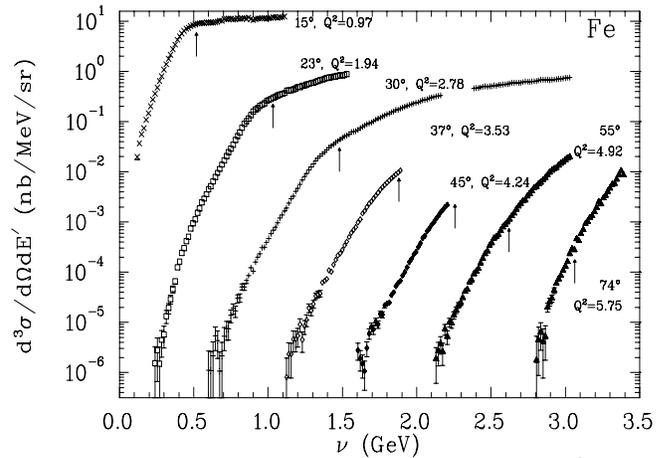}
\caption{Differential cross section for Fe. The $Q^2$ values given at each 
angle correspond to Bjorken $x = 1$. The value of $\nu$ for $x = 1$ is
shown by an arrow for each kinematic setting.
Statistical errors only are shown.}
\label{sigma}
\end{figure}

Fig.~\ref{sigma} shows the measured cross sections vs. energy loss $\nu$ 
for Fe, where for each angle the $Q^2$ value at Bjorken
$x = Q^2/2M\nu = 1$ is given (this value corresponds to elastic 
scattering from a free nucleon). 
Because of the significant smearing due to the
Fermi motion and the large contribution from other inelastic processes 
(e.g. $\pi$ production, resonance production and deep inelastic scattering) at 
these relatively high $Q^2$, there is little evidence of a quasielastic peak. 
In fact the sharp bend in the spectrum at $\theta = 15^\circ$ is the only 
distinctive feature resulting from quasielastic scattering. At larger
angles the additional inelastic processes cause even this feature to
disappear. It should be noted however that quasielastic scattering
is still expected to contribute significantly to the cross section for
$\nu < Q^2/2M$ ($x > 1$). The minimum measured cross sections 
were limited by count rate and represent a factor of $ > 100$ improvement
in sensitivity compared to the previous experiment~\cite{ne3}. This 
improvement is largely due to the higher beam currents and larger 
acceptance spectrometers available at TJNAF. 

The scaling function is defined as the ratio of the measured cross section 
to the off-shell electron-nucleon cross section multiplied by a kinematic
factor: 
$$
F(y) = {d^2\sigma \over d\Omega d\nu}[Z \sigma_p + N \sigma_n]^{-1}{q \over (M^2 + (y + q)^2)^\frac{1}{2}}
$$
Where Z and N are the number of protons and neutrons in the target nucleus,
the off-shell cross sections $\sigma_p$ and $\sigma_n$ are taken from 
$\sigma_{CC1}$ from Ref.~\cite{deforest} using the elastic 
form factors from Ref.~\cite{gari},
$q$ is the three-momentum transfer and M is the mass of the proton. 

The $y$ variable is defined through the equation~\cite{pacesalme}:
$$
\nu + M_A = (M^2 + q^2 + y^2 + 2yq)^{\frac{1}{2}} + (M_{A-1}^2 + y^2)^\frac{1}{2}
$$
where $M_A$ is the mass of the target nucleus and $M_{A-1}$ is the ground 
state mass of the $A-1$ nucleus. 

\begin{figure}[yfe]
\epsfxsize 8.5 cm \epsfysize 6 cm \epsfbox{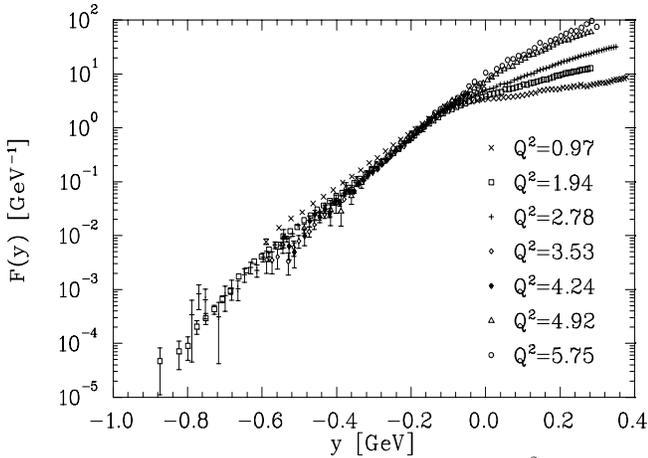}
\caption{Scaling function $F(y)$ for Fe. The $Q^2$ values are given for 
Bjorken $x = 1$}
\label{yfe}
\end{figure}

The scaling function for Fe is shown in Fig.~\ref{yfe} for all measured 
angles. While the cross section as a function of $Q^2$ and $\nu$ varies
over many orders of magnitude (see Fig. 1), the scaling function for 
values of $y < -0.1$ GeV/c shows a clear approach to a universal curve
where the data can be represented by a function that depends only on $y$. 
The breakdown of scaling for values of $y$ near zero is due to the
dominance of other inelastic processes beyond quasielastic scattering. 

\begin{figure}[yscale1]
\epsfxsize 8.5 cm \epsfbox{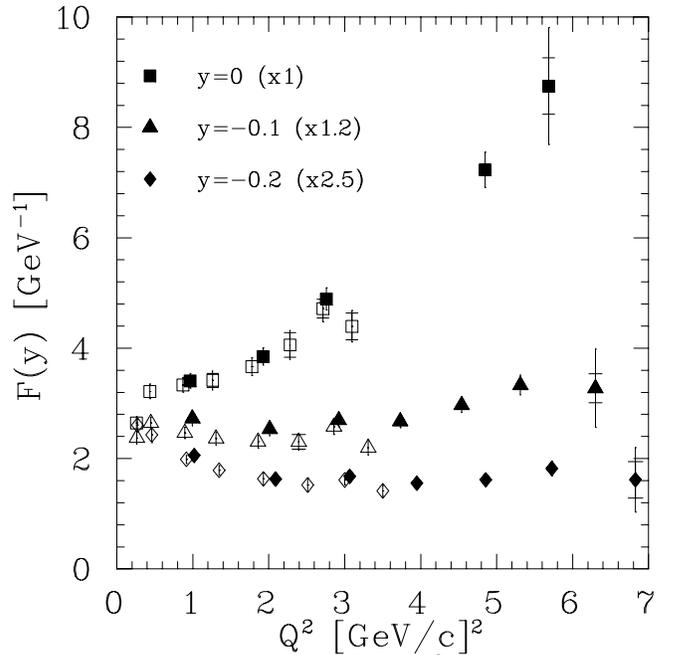}
\caption{Scaling function $F(y)$
vs. $Q^2$ for Fe for fixed values of $y = 0, -0.1, -0.2$ GeV/c. 
The open points are calculated from the measured cross sections of
Ref. [15] including Coulomb corrections and using the definition of 
$y$ as discussed in the text. 
The scaling functions for each value of $y$ have been multiplied by the 
factors in parentheses. The inner error bar is the statistical 
uncertainty and the outer error bar is the statistical and
systematic uncertainty added in quadrature.}

\label{yscale1}
\end{figure}

\begin{figure}[yscale2]
\epsfxsize 8.5 cm \epsfbox{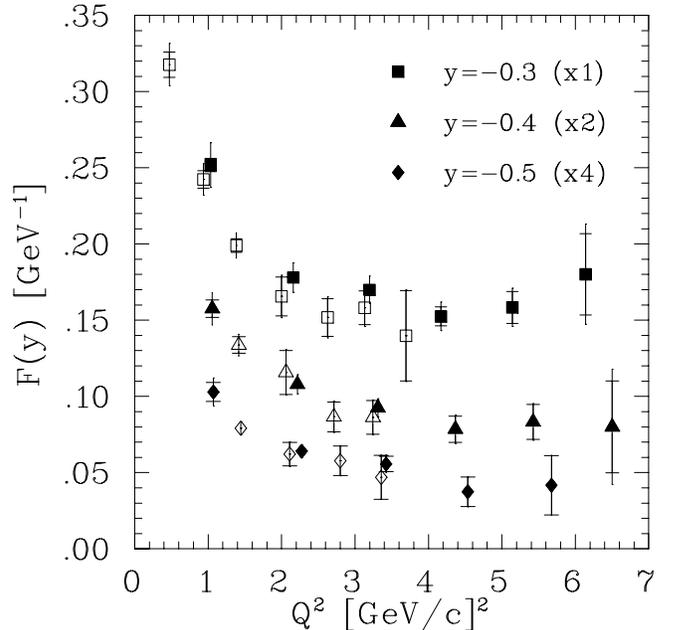}
\caption{Same as Fig.~\ref{yscale1} for fixed 
values of $y = -0.3, -0.4, -0.5$ GeV/c.}
\label{yscale2}
\end{figure}

The approach to scaling is also shown in Figs.~\ref{yscale1} and \ref{yscale2},
where the $Q^2$ dependence of $F(y)$ at several fixed values of $y$ is 
presented. For $y = -0.2$ to $-0.5$ GeV/c there is a clear approach to scaling
as $Q^2$ is increased. This is the first evidence for $y$-scaling in heavy 
nuclei for $y < -0.3$ GeV/c. 
There are, in addition, significant scaling violations
observed at both low and high $Q^2$. The increase in $F(y)$ with $Q^2$ for 
$y = 0, -0.1$ GeV/c (Fig.~\ref{yscale1}) 
is clearly due to the inelastic processes mentioned above. A similar effect
was observed~\cite{ne18} previously, but only for $y \sim 0$. 
Calculations that include both quasielastic and other inelastic 
processes~\cite{ben2,fsi8} 
indicate that at $y = 0$ these other process dominate the reaction for 
$Q^2 > 2$ (GeV/c)$^2$. 

At large negative $y$ (Fig.~\ref{yscale2}) 
there is a decrease in $F(y)$ with increasing $Q^2$ as the 
scaling is approached. This behavior contradicts the approach to scaling
expected within the impulse approximation (where the scaling limit is
approached from below because of incomplete
kinematic coverage at low $Q^2$), and suggests the influence of
final state interactions. A recent calculation~\cite{simpriv} indicates 
that the
component of the FSI resulting from the scattered nucleon interacting with
the mean-field of the nucleus should be a strongly decreasing function of 
$Q^2$ and 
become negligible for $Q^2 > 3$ (GeV/c)$^2$. 
An additional component in the calculation, due to interaction with a 
correlated nucleon,
has a much weaker $Q^2$ dependence and may persist to the $Q^2$ range
of the present experiment. The present data suggest a scaling that is
consistent with an approach to the impulse approximation scaling limit,
but cannot exclude contributions from FSI that are 
$Q^2$-independent. 

\begin{figure}[adep]
\epsfxsize 7.5 cm \epsfbox{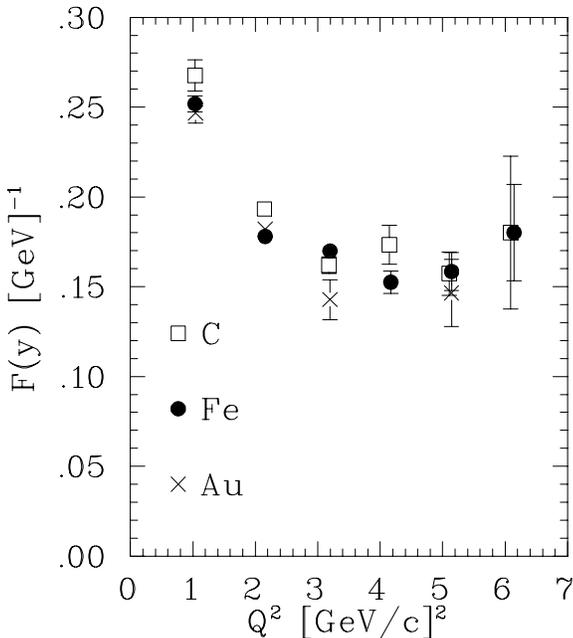}
\caption{Scaling function vs. $Q^2$ for C, Fe and Au at $y = -0.3 ~GeV/c$.
Error bars are statistical only. 
} 
\label{adep}
\end{figure}

Comparison of the scaling functions for C, Fe and Au show very similar
distributions. This can be seen in Fig.~\ref{adep}, where all targets
are plotted vs. $Q^2$ for a fixed value of $y = - 0.3$ GeV/c. The small
A-dependence seen in these data is suggestive of a universal response for
all medium-mass nuclei as might be expected in a kinematic region 
dominated by short-range correlations. 

In summary, we have measured the inclusive cross section at $x > 1$
for electrons scattering from C, Fe and Au targets to $Q^2 \simeq 7$ 
(GeV/c)$^2$, a significant increase compared to the previous experiment.
When analyzed in terms of the $y$-scaling function the data show an 
approach to scaling for $Q^2 > 3$ (GeV/c)$^2$. At these values of $Q^2$ a 
scaling limit can be expected within a simple impulse approximation.
In addition a scaling behavior is observed for the first time
at very large negative $y$ ($y = -0.5$ GeV/c). 
This is a regime where the nucleon momentum distribution 
is expected to be dominated by short-range nucleon-nucleon correlations.
It is interesting to note that
contributions from short-range final-state interactions  
may also result in a scaling-like behavior due to the small $Q^2$-dependence 
of these effects, and that these contributions are also dominated by 
short-range nucleon-nucleon correlations.

We gratefully acknowledge the staff and management of TJNAF for
their efforts in delivering the electron beam. We also acknowledge helpful 
discussions with O. Benhar and C. Ciofi degli Atti.
This research was supported by the National Science Foundation, the
Department of Energy and the Swiss National Science Foundation. 

$^*$ Present Address: Physics Division, Argonne National Laboratory, Argonne, IL 60439.

$^\#$ Present Address: Thomas Jefferson National Accelerator Facility, Newport News, VA 23606.

$^+$ Present Address: College of William and Mary, Williamsburg, VA 23187.

$^\&$ Present Address: Mississippi State University, Mississippi State, MS 39762.



\end{document}